\documentstyle[twoside,epsfig,fleqn,espcrc2]{article}

\newcommand{\AmS}{{\protect\the\textfont2
  A\kern-.1667em\lower.5ex\hbox{M}\kern-.125emS}}
\newcommand\sfrac[2]{\!\! \begin{array}{c} \frac{#1}{#2} \end{array} \!\!}  

\newcommand{\mIm}{\,\mbox{\small $\Im$m\,}} 
\newcommand{\eRe}{\,\mbox{\small $\Re$e\,}} 
\newcommand{\be}{\begin{eqnarray}} 
\newcommand{\ee}{\end{eqnarray}} 
\newcommand{\nn}{\nonumber} 
\newcommand{\dgs}{d^{\gamma}{\!\scriptstyle (}s{\scriptstyle )}\,} 
\newcommand{\dzs}{d^{Z}{\!\scriptstyle (}s{\scriptstyle )}\,}

\title{Electroweak dipole moment form factors
       of the top quark in supersymmetry}

\author{A. Bartl\address{Institut 
         f\"ur Theoretische Physik, 
         Universit\"at Wien, \\ A-1090 Vienna, Austria
        }\thanks{Talk presented at the XVI International
                 Workshop on Weak Interaction and Neutrinos (WIN 97), 
                 Capri, Italy, 22-28 June, 1997. 
                 To appear in the Proceedings},
        E. Christova\address{Institute of Nuclear Research and Nuclear Energy, 
         \\ Boul. Tzarigradsko Chaussee 72, Sofia 1784, Bulgaria},
        T. Gajdosik{\hbox{$^{\rm c}$}},
        W. Majerotto\address{Institut f\"ur Hochenergiephysik der 
         \"Osterreichischen Akademie der Wissenschaften, \\
         A-1050 Vienna, Austria}}

\begin{document} 

\begin{abstract}

\begin{picture}(0,0)
\put(445,270){\makebox(0,0)[ r]{HEPHY-PUB 672/97}}
\put(445,260){\makebox(0,0)[ r]{UWThPh-1997-28}}
\put(445,250){\makebox(0,0)[ r]{hep-ph/9709219}}
\end{picture}
We present a 
complete analysis of the electrtic and weak dipole moment form 
factors of the top quark within the 
Minimal Supersymmetric Standard Model with complex parameters. We 
include gluino, chargino, and neutralino exchange in the loops of the 
\mbox{$\gamma t\bar{t}$} and \mbox{$Z t\bar{t}$} vertices.

\end{abstract}

\maketitle

\section{Introduction}
The large mass of the top quark 
allows one to probe physics at a high energy scale, where 
new physics might show up. In the last years a number of 
papers~\cite{{CPviol},{dipgl},{we}} 
considered CP violating 
observables in top quark production as tests for new physics. In   
$e^{+} e^{-}$ annihilation these effects are due to the weak and 
electric dipole moment form factors $\dzs$ and $\dgs$ of the top quark.
In general, the $\gamma t \bar{t}$ and $Z t\bar{t}$ vertices including 
the CP violating form factors are  
\be 
e ( {\cal V}_{\gamma}^{t} )_{\mu}  
&=& e \Bigl( \frac{2}{3} \gamma_{\mu} 
 - i \, \frac{\dgs}{m_{t}} {\cal P}_{\mu} \gamma_{5} \Bigr) 
\, , 
\label{photonvertex} 
\\  
g_{Z} ( {\cal V}_{Z}^{t} )_{\mu}  
&=& g_{Z}
\Bigl( \gamma_{\mu} ( g_{V} + g_{A} \gamma_{5} )  
\nn\\ & & \quad
 - i \frac{\dzs}{m_{t}} {\cal P}_{\mu} \gamma_{5} \Bigr) 
\, ,
\label{Zvertex} 
\ee 
where $\dzs$ and $\dgs$ are the weak and electric dipole moment form 
factors of the top quark, and 
\mbox{$g_{V} = (1/2) - (4/3) \sin^{2}\Theta_{W}$},
\mbox{$g_{A} = - (1/2)$}, 
\mbox{${\cal P}_{\mu} = p_{t\,\mu} - p_{\bar{t}\,\mu}$}, 
\mbox{$g_{Z} = g / (2 \cos \Theta_{W}$}, 
and 
\mbox{$g = e/\sin\Theta_{W}$} with $e$ the electro--magnetic 
coupling constant and $\Theta_{W}$ the 
Weinberg angle.
 
In the Standard Model (SM) CP violation can  
appear only through the phase in the CKM--matrix.  
The dipole moment form factors $\dgs$ and $\dzs$ for the quarks  
are at least two--loop order effects and hence very small.  
In the Minimal  
Supersymmetric Standard Model (MSSM)~\cite{Kane} additional  
complex couplings can be introduced that lead to  
CP violation within one generation only~\cite{Dugan}, and which occur 
at one--loop level. If the masses 
of the SUSY particles are not very much higher than the mass  
of the top, one expects SUSY radiative corrections to  
induce larger values for $\dgs$ and $\dzs$.  
According to the particle content in  
the loop (see Fig. 1a,b) we distinguish the following three 
contributions: 
\begin{figure*}
\begin{center}
\setlength{\unitlength}{1mm}
%
%
\begin{picture}(150,50)(0,15)
\put(  0, 5){\mbox{\epsfig{file=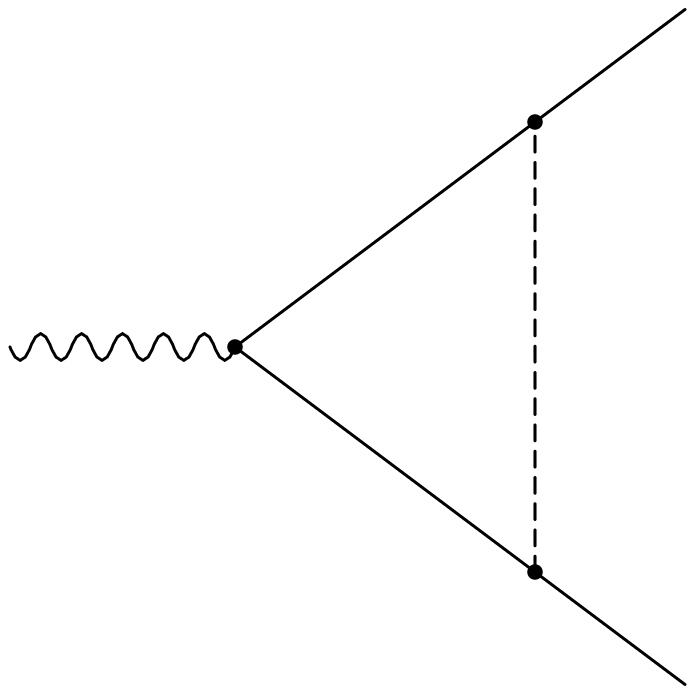,width=6cm}}}
\put( 10,37){\makebox(0,0)[b ]{$A^{\mu}, Z^{\mu}$}}
\put( 60,62){\makebox(0,0)[tl]{$\bar{t}$}}
\put( 60, 6){\makebox(0,0)[bl]{$t$}}
\put( 35,46){\makebox(0,0)[br]{$\overline{\tilde{\chi}}^{0}_{k}, 
                                \overline{\tilde{\chi}}^{+}_{k}$}}
\put( 35,24){\makebox(0,0)[tr]{$\tilde{\chi}^{0}_{j}, 
                                \tilde{\chi}^{+}_{j}$}}
\put( 47,34){\makebox(0,0)[ l]{$\begin{array}{l}
                                  \tilde{t}_{m} \, , \\ 
                                  \tilde{b}_{m}
                                \end{array}$}}
\put( 30, 5){\makebox(0,0)[b ]{{\large\bf Fig.~1a}}}
\put( 70, 5){\mbox{\epsfig{file=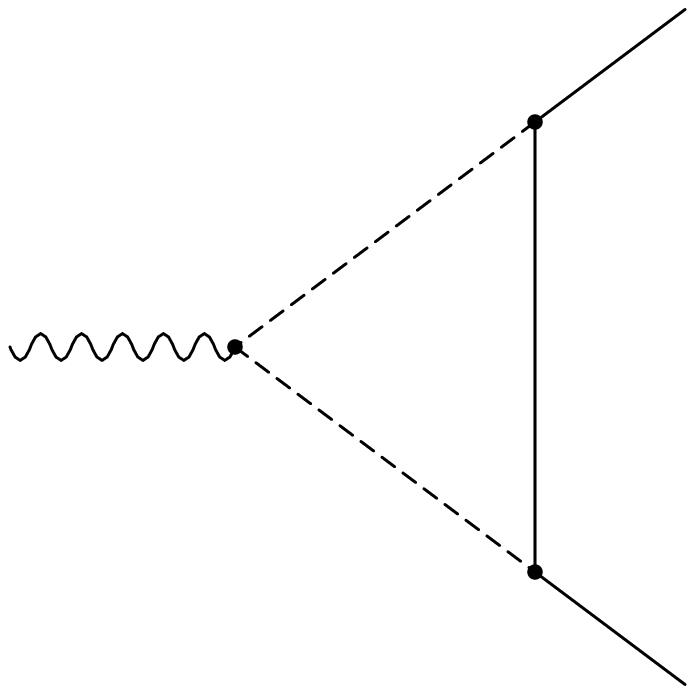,width=6cm}}}
\put( 80,37){\makebox(0,0)[b ]{$A^{\mu}, Z^{\mu}$}}
\put(130,62){\makebox(0,0)[tl]{$\bar{t}$}}
\put(130, 6){\makebox(0,0)[bl]{$t$}}
\put(105,46){\makebox(0,0)[br]{$\tilde{t}_{m}^{*}, \tilde{b}_{m}^{*}$}}
\put(105,24){\makebox(0,0)[tr]{$\tilde{t}_{n}, \tilde{b}_{n}$}}
\put(117,34){\makebox(0,0)[ l]{$\begin{array}{l}
                                  \tilde{g} \, , \\ 
                                  \tilde{\chi}^{0}_{k} \, , \\ 
                                  \tilde{\chi}^{+}_{k}
                                \end{array}$}}
\put(100, 5){\makebox(0,0)[b ]{{\large\bf Fig.~1b}}}
\end{picture}\\
\end{center}
\caption{Feynman diagrams contributing to $\dgs$ and $\dzs$: }
   {\bf (a)} with two fermions and one scalar in the loop, 
   {\bf (b)} with two scalars and one fermion in the loop.
\setlength{\unitlength}{1pt}
\end{figure*}
\begin{enumerate} 
  \item  
    gluino contribution $d^{\gamma, Z}_{\tilde{g}}$ with  
    (\,$\tilde{t}\,\bar{\tilde{t}}\,\tilde{g}$\,) in the loop, 
  \item  
    chargino contribution $d^{\gamma, Z}_{\tilde{\chi}^{+}}$ 
    with (\,$\tilde{\chi}^{+}\,\tilde{\chi}^{-}\,\tilde{b}$\,) 
    and (\,$\tilde{b}\,\bar{\tilde{b}}\,\tilde{\chi}^{+}$\,) in the loop, 
  \item  
    neutralino contribution $d^{\gamma, Z}_{\tilde{\chi}^{0}}$ 
    with (\,$\tilde{\chi}^{0}\,\tilde{\chi}^{0}\,\tilde{t}$\,) 
    and (\,$\tilde{t}\,\bar{\tilde{t}}\,\tilde{\chi}^{0}$\,) in the loop. 
\end{enumerate} 
 
The gluino contribution $d^{\gamma, Z}_{\tilde{g}}$ was considered in  
\cite{we}. The chargino and 
neutralino contributions were calculated in \cite{bcgm}. 
Although the gluino contribution  
$d^{\gamma, Z}_{\tilde{g}}$ is proportional to $\alpha_{s}$ it turns 
out that the chargino contribution  
$d^{\gamma, Z}_{\tilde{\chi}^{+}}$, which is  
proportional to $\alpha_{w}$,  
\mbox{($\alpha_{w} = g^{2}/(4\pi)$)}  
can be equally important (see also \cite{Osh1}). This is due to  
threshold enhancements which can occur in the contributions of the 
diagrams in Fig. 1a, and the large Yukawa couplings: 
\mbox{$Y_{t} = m_{t}/(\sqrt{2} m_{W} \sin\beta)$} and 
\mbox{$Y_{b} = m_{b}/(\sqrt{2} m_{W} \cos\beta)$}.  
In general the neutralino contribution turns out to be smaller. 
However, there are cases where it is important. 

\section{Complex couplings in the MSSM}
In the MSSM the Higgs--higgsino mass parameter $\mu$ and the trilinear 
scalar coupling parameters $A_{t}$ and $A_{b}$ can be complex, and 
provide the CP violating phases. The 
calculation requires the diagonalization of the squark, chargino, and 
neutralino mass matrices.
We use the singular value decomposition \cite{SingW} to diagonalize 
the complex neutralino and chargino mass matrices. 
 
The size of the dipole moment form factors $\dgs$, $\dzs$  
depends strongly on the phases of the SUSY parameters. 
There are constraints \cite{{Osh1},{Garisto}} on some phases  
from the measurement of the electric dipole moment (EDM) 
of the neutron. Usually, 
one concludes~\cite{{Osh1},{Garisto}} 
that either the phases involved in the EDM of the neutron are  
very small or the masses of the first generation of squarks 
are in the TeV range. By using supergravity (SUGRA) with grand 
unification (GUT) there are attempts 
to constrain also the phases entering the dipole moments of the  
top. In our analysis we want to be more 
general and we do not make any additional assumptions about GUT 
except for the unification of the gauge couplings and gaugino masses. 
In particular, 
we do not assume unification of the scalar mass parameters and of 
the trilinear scalar coupling parameters $A_{q}$  
of the different generations.

In the MSSM with complex phases, $\dgs$ and $\dzs$ are generated in 
one--loop order, irrespectively of generation mixing. 
We treat the chargino and neutralino contributions
separately from the gluino contribution 
not only because different couplings are involved (electroweak 
and strong), but also because they are sensitive to different 
SUSY parameters. The contributions from the different Feynman 
diagrams (Fig.~1a,b) depend in a distinctive way on the
SUSY parameters. 
The Passarino--Veltman three point functions 
$C_{0}$, $C_{i}$, and $C_{ii}$ ($i = 1,2$)~\cite{PaVe} appear in 
the loop integrations of the diagrams Fig.~1a,b. 

\section{Numerical results}
The chargino contribution  depends 
on the gaugino and higgsino couplings, as well as  on  
the squark mixing angle and phase. 
We have included the terms proportional to the bottom Yukawa 
coupling $Y_{b}$ which are important for large $\tan\beta$. For small 
values of $\tan\beta$ the terms proportional to $Y_{t}$ dominate.
As  neutralinos do not couple to the photon, 
$d^{\gamma}_{\tilde{\chi}^{0}}$ receives a non--zero 
contribution only 
from the diagram with $\tilde{t}\,\bar{\tilde{t}}\,\tilde{\chi}^{0}$ 
exchanged in the loop.
\begin{figure*}
\begin{center}
\setlength{\unitlength}{1mm}
%
%
\begin{picture}(150,65)(0,15)
\put( 7,-5.5){\mbox{\epsfig{file=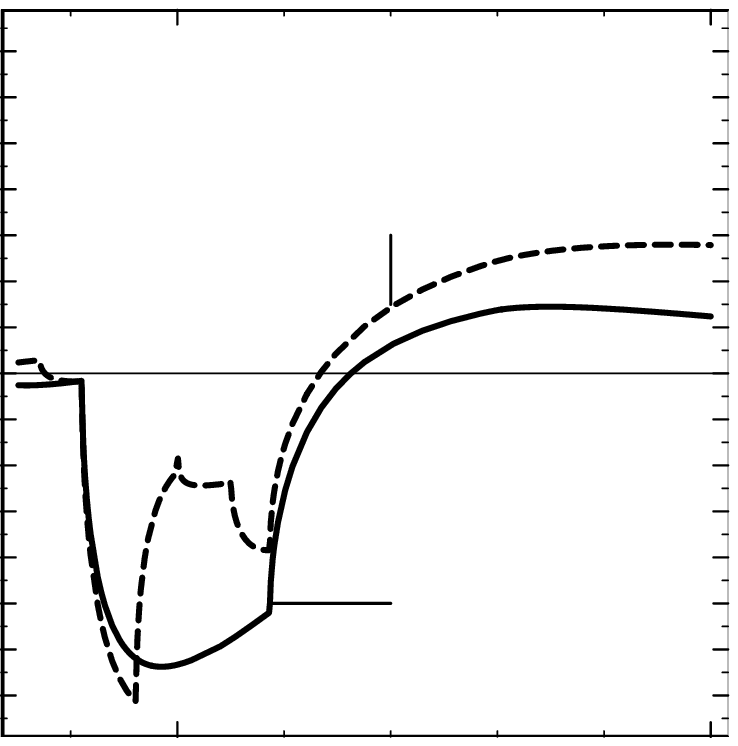,width=63mm}}}
\put( 41,59 ){\makebox(0,0)[b ]{$\mIm \dzs$}}
\put( 42,27 ){\makebox(0,0)[ l]{$\mIm \dgs$}}
\put(  6,71){\makebox(0,0)[ r]{\small  0.0006}}
\put(  6,63){\makebox(0,0)[ r]{\small  0.0004}}
\put(  6,55){\makebox(0,0)[ r]{\small  0.0002}}
\put(  6,47){\makebox(0,0)[ r]{\small  0}}
\put(  6,39){\makebox(0,0)[ r]{\small -0.0002}}
\put(  6,31){\makebox(0,0)[ r]{\small -0.0004}}
\put(  6,23){\makebox(0,0)[ r]{\small -0.0006}}
\put( 13  ,14){\makebox(0,0)[t]{\small 400}}
\put( 22.2,14){\makebox(0,0)[t]{\small 500}}
\put( 31.4,14){\makebox(0,0)[t]{\small 600}}
\put( 40.6,14){\makebox(0,0)[t]{\small 700}}
\put( 49.8,14){\makebox(0,0)[t]{\small 800}}
\put( 59  ,14){\makebox(0,0)[t]{\small 900}}
\put( 70, 5){\makebox(0,0)[b ]{$\sqrt{s}/$GeV}}
\put( 38, 4){\makebox(0,0)[b ]{\large\bf Fig.~2a}}
\put(70,-5.5){\mbox{\epsfig{file=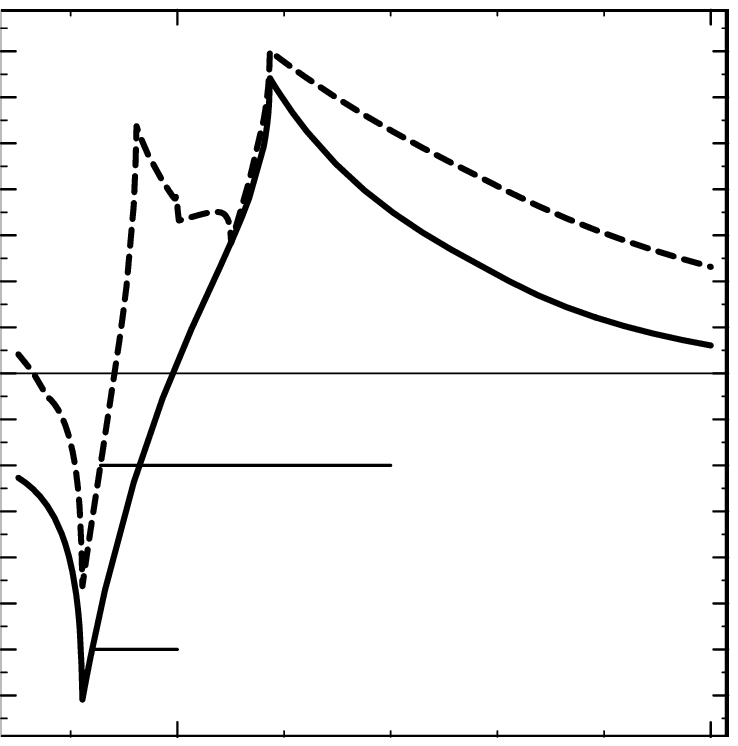,width=63mm}}}
\put(104,39){\makebox(0,0)[ l]{$\eRe \dzs$}}
\put( 86,23){\makebox(0,0)[ l]{$\eRe \dgs$}}
\put(134,71){\makebox(0,0)[ l]{\small  0.0006}}
\put(134,63){\makebox(0,0)[ l]{\small  0.0004}}
\put(134,55){\makebox(0,0)[ l]{\small  0.0002}}
\put(134,47){\makebox(0,0)[ l]{\small  0}}
\put(134,39){\makebox(0,0)[ l]{\small -0.0002}}
\put(134,31){\makebox(0,0)[ l]{\small -0.0004}}
\put(134,23){\makebox(0,0)[ l]{\small -0.0006}}
\put( 76  ,14){\makebox(0,0)[t]{\small 400}}
\put( 86.2,14){\makebox(0,0)[t]{\small 500}}
\put( 94.4,14){\makebox(0,0)[t]{\small 600}}
\put(103.6,14){\makebox(0,0)[t]{\small 700}}
\put(112.8,14){\makebox(0,0)[t]{\small 800}}
\put(122  ,14){\makebox(0,0)[t]{\small 900}}
\put(101, 4){\makebox(0,0)[b ]{\large\bf Fig.~2b}}
\end{picture}\\
\setlength{\unitlength}{1pt}
\end{center}
\caption{
  $\dgs$ and $\dzs$ for the reference parameter set with 
  $M = 230$~GeV: }
  {\bf (a)} 
  $\mIm \dgs$ (full line), $\mIm \dzs$ (dashed line) and
  {\bf (b)}
  $\eRe \dgs$ (full line), $\eRe \dzs$ (dashed line).
\end{figure*}

In the following we give numerical results for the real and 
imaginary parts of $\dgs$ and $\dzs$. 
Quite generally they depend on the parameters 
$M^{\prime}$, $M$, $|\mu|$, $\tan\beta$, $m_{\tilde{t}_{k}}$,
$m_{\tilde{b}_{k}}$, $\cos\theta_{\tilde{t}}$, 
$\cos\theta_{\tilde{b}}$ and the phases $\varphi_{\mu}$, 
$\varphi_{\tilde{t}}$, $\varphi_{\tilde{b}}$,  and
$\varphi_{\tilde{g}}$.  
The GUT relations
\be 
m_{\tilde{g}} &=& (\alpha_{s}/\alpha_{2}) M \approx 3 M
\\
M^{\prime} &=& \sfrac{5}{3} \tan^{2}\Theta_{W} M
\ee
imply that the gaugino mass parameters have the same phase. 

We take \mbox{$m_{W} = 80$~GeV}, \mbox{$m_{t} = 175$~GeV}, 
\mbox{$m_{b} = 5$~GeV}, \mbox{$\sqrt{s} = 500$~GeV}, 
\mbox{$\alpha_{s} = 0.1$}, and \mbox{$\alpha_{em} = \sfrac{1}{123}$}.
In order not to vary too many parameters we choose the following set of
SUSY parameter values:
  $M = 230$~GeV,
    $m_{\tilde{t}_{1}}=150$~GeV,
      $m_{\tilde{b}_{1}}=270$~GeV,
  $|\mu| = 250$~GeV, 
    $m_{\tilde{t}_{2}}=400$~GeV, 
      $m_{\tilde{b}_{2}} = 280$~GeV, 
  $\tan\beta = 2$,
    $\theta_{\tilde{t}}=\frac{\pi}{9}$,
      $\theta_{\tilde{b}}=\frac{\pi}{36}$,
  $\varphi_{\mu}=\frac{4 \pi}{3}$,
    $\varphi_{\tilde{t}}=\frac{\pi}{6}$,
      $\varphi_{\tilde{b}}=\frac{\pi}{3}$. 

Notice that the dipole moment form factors $\dgs$ and $\dzs$ depend 
on $\varphi_{\mu}$ not only through the chargino and neutralino 
diagonalizing matrices, but also 
through the chargino and neutralino mass spectra. The 
values of the chargino and neutralino masses can vary by about 
40 percent when \mbox{$\cos\varphi_{\mu}$} is varied between $-1$ 
and $1$. 
 
\mbox{$\mIm \dgs$} and \mbox{$\mIm \dzs$} are
determined by the absorptive parts of the amplitudes. Therefore they 
vanish when no real production of charginos or neutralinos is possible.
Local maxima occur near the thresholds of chargino or neutralino 
pair production,
and they get bigger if the gaugino and higgsino component of the 
chargino are approximately equal.
The neutralino contribution to \mbox{$\mIm \dgs$} 
and \mbox{$\mIm \dzs$}
is one order of magnitude 
smaller than the chargino contribution because the photon does not 
couple to the neutralinos and only Fig.~1b contributes. 
The neutralino contribution \mbox{$\mIm d^{Z}_{\tilde{\chi}^{0}}$} 
is smaller than the chargino 
contribution because the couplings are smaller. It 
shows the same qualitative behaviour as the chargino contributions,
but it is more complicated because of the richer particle spectrum. 
Note that the two neutralinos in Fig.~1a have to be different.   

In Fig.~2a we show \mbox{$\mIm \dgs$} and \mbox{$\mIm \dzs$}
as functions of $\sqrt{s}$, where 
all contributions (gluino, chargino, and neutralino) are summed up. 
The threshold effects can be seen very clearly.
There is a big enhancement in $\dgs$ and $\dzs$ because the 
threshold for $\tilde{\chi}^{+}_{1}\tilde{\chi}^{-}_{1}$ production 
is reached at \mbox{$\sqrt{s}=420$~GeV} for 
\mbox{$m_{\tilde{\chi}^{+}_{1}} = 210$~GeV}. At 
\mbox{$\sqrt{s}=590$~GeV} $\tilde{\chi}^{+}_{2}\tilde{\chi}^{-}_{2}$ 
production becomes possible and again there is a big contribution but 
with the opposite sign. The additional thresholds in $\dzs$ are due to 
the neutralino contributions. In Fig.~2b we show 
 \mbox{$\eRe \dgs$} and \mbox{$\eRe \dzs$}, which 
can also be understood via 
dispersion relations: each spike corresponds to the opening of a new 
production channel.

\bigskip

\section*{Acknowledgements}

We thank Helmut Eberl for his constructive assisitance in the 
evaluation of the loop integrals and Sabine Kraml for the discussions 
about the couplings in the Lagrangian. 
We also thank Stefano Rigolin for his helpful correspondence 
regarding the numerical calculations. 
E.C.'s work has been 
supported by the Bulgarian National Science Foundation, Grant Ph--510. 
This work was also supported by the 'Fonds zur F\"orderung der 
wissenschaftlichen Forschung' of Austria, project no. P10843--PHY. 

\bigskip

\begin{flushleft}

\end{flushleft}

\end{document}